\documentclass[%
reprint,
superscriptaddress,
nofootinbib,
amsmath,
amssymb,
aps,
floatfix,
showkeys,
]{revtex4-2}

\usepackage[colorlinks,citecolor=blue,linkcolor=blue,anchorcolor=blue,filecolor=blue,
urlcolor=blue]{hyperref}

\usepackage{graphicx,color,xcolor}
\usepackage{amsfonts,amsmath}
\usepackage{amssymb,ulem}
\usepackage{dcolumn}
\usepackage{bm,lipsum}
\usepackage[utf8]{inputenc}
\usepackage{subeqnarray}
\usepackage{array}
\usepackage{epsfig,hyperref}
\usepackage{morefloats}
\usepackage{relsize}
\usepackage{url}

\usepackage{comment}

\begin{document}

\preprint{ }

\title{Imprints of the nuclear symmetry energy slope in gravitational wave signals emanating from neutron stars}

\author{Luiz L. Lopes}
\email{llopes@cefetmg.br}
\affiliation{Centro Federal de Educa\c{c}\~ao Tecnol\'ogica de Minas Gerais Campus VIII, Varginha/MG, CEP 37.022-560, Brazil}

\author{Victor B. T. Alves}
\affiliation{Departamento de F\'isica, CCET - Universidade Federal do Maranh\~ao, Campus Universit\'ario do Bacanga; CEP 65080-805, S\~ao Lu\'is, MA, Brasil.}

\author{César O. V. Flores}
\email{cesar.vasquez@uemasul.edu.br}
\affiliation{Centro de Ci\^encias Exatas, Naturais e Tecnol\'ogicas, CCENT - Universidade Estadual da Regi\~ao Tocantina do Maranh\~ao; C.P. 1300,\\ CEP 65901-480, Imperatriz, MA, Brasil.}
\affiliation{Departamento de F\'isica, CCET - Universidade Federal do Maranh\~ao, Campus Universit\'ario do Bacanga; CEP 65080-805, S\~ao Lu\'is, MA, Brasil.}

\author{Germán Lugones}
\email{german.lugones@ufabc.edu.br}
\affiliation{Universidade Federal do ABC, Centro de Ci\^encias Naturais e Humanas, Avenida dos Estados 5001 - Bang\'u, CEP 09210-580, Santo Andr\'e, SP, Brazil.}

\date{\today}

\begin{abstract}
We investigate possible traces of the nuclear symmetry energy slope ($L$) in the gravitational wave emission of neutron stars. For fixed stellar mass values, we examine how the slope influences the stellar radius, compactness, the tidal deformability, the frequency of the quadrupole fundamental fluid mode, and the damping time of the mode due to the gravitational wave emission. We demonstrate that all these physical quantities are sensitive to the slope and could potentially impose significant constraints on it.
\end{abstract}


\maketitle

\section{Introduction}
Our knowledge of nuclear physics has taken a great leap forward in the last decade. However, as of today, there is a giant fog blurring our vision of a crucial nuclear property: the slope of the symmetry energy at the saturation point ($L$). 
In the early 2010s, most studies pointed to a relatively low value of $L$. For instance, references~\cite{Paar2014,Lattimer2013,Steiner2014} presented upper limits of 54.6, 61.9, and 66 $\mathrm{MeV}$, respectively. However, in the last couple of years, much higher upper limits have appeared in the literature. For example, a study of the spectra of pions in intermediate energy collisions pointed to an upper limit of 117.5 $\mathrm{MeV}$\cite{pions}, while one of the PREX2 analyses \cite{PREX2} suggested even larger values. On the other hand, in a recent paper~\cite{Tagami2022}  a potential conflict among these results was highlighted:  the CREX group reported a slope in the range $0 < L (\mathrm{MeV})  < 51$, while the PREX2 results pointed to $76 < L (\mathrm{MeV}) < 165$; both at the 68\% confidence level. Although there is overlapping between them at the 90\% confidence level \cite{Zhang:2022bni}, the PREX results are in tension with CREX, as well as with predictions of chiral effective field theory \cite{PREX2}.

Due to the potentially conflicting results obtained in terrestrial laboratories, we turn our attention to studying neutron stars, which may provide valuable insights.
Just like in nuclear physics, our knowledge of neutron stars has dramatically increased in the last decade. We can highlight the discovery of two-solar masses pulsars, such as PSR J0348+0432~\cite{Antoniadis}, as well as results from the NICER X-ray telescope~\cite{Riley:2019yda,Miller:2019cac} and the LIGO/VIRGO gravitational wave observatories~\cite{Abbott2017,Abbott:2018wiz,AbbottPRL}.

In this work, we study how the symmetry energy slope $L$ affects some neutron star properties. To accomplish this task, we use quantum hadrodynamics (QHD) with the traditional $\sigma-\omega-\rho$ mesons~\cite{Serot_1992}. Moreover, to keep the symmetry energy fixed while varying the slope, we use two extensions of QHD: to reduce the slope, we add the non-linear $\omega-\rho$ coupling as presented in the IUFSU model~\cite{IUFSU,Rafa2011,dex19jpg}, while to increase the slope, we add the scalar-isovector $\delta$ meson~\cite{KUBIS1997,Liu2002,Lopes2014BJP}. Using two different interactions allows us not only to understand the phenomenological point of view about the influence of changes in $L$, but also the field theory point of view about the influence of different fields and couplings.

We start by studying how the slope, $L$, affects the classical, macroscopic mass-radius relation~\cite{Rafa2011,Bao2020PRC}. Although we demonstrate how the slope affects the entire neutron star family, we also focus explicitly on two specific values: the 1.4 $M_\odot$ star, which is called the canonical star, and the 2.01 $M_\odot$ star, which is not only the most probable mass value of PSR J0348+0432~\cite{Antoniadis}, but also the lower limit of PSR J0740+6620, whose gravitational mass is 2.08 $\pm$ 0.07 $M_\odot$~\cite{Miller2021,Riley2021}. Therefore, any equation of state (EOS) unable to reach at least 2.01 $M_\odot$ must be ruled out.

We then analyze the influence of the slope on the dimensionless tidal parameter $\Lambda$ \cite{Bao2020PRC}. The tidal deformability of a compact object is related to how easily the object is deformed when subjected to an external tidal field. A larger tidal deformability indicates that the object is easily deformable. Conversely, a compact object with a small tidal deformability parameter is more compact and more difficult to deform \cite{Chat2020}. Binary neutron star mergers, such as GW170817 detected by the LIGO/VIRGO observatories~\cite{Abbott:2018wiz,Abbott2017,AbbottPRL}, provide us with additional information about the neutron star's EOS. In these types of events, the neutron star components begin to react to their mutual tidal fields before the merging, and this effect can be detected in the phase modification of the gravitational wave impinging on the detector. This tidal response strongly depends on the neutron star EOS, and therefore important information can be obtained about it~\cite{Eanna2008,Read2009,Flores2020}.

Finally, neutron star oscillations can provide valuable information about the microphysical properties of dense matter. A wide variety of pulsation modes can be excited in newly born compact objects, such as those associated with the violent dynamics of core-collapse supernovae~\cite{Camelio:2017nka}, starquakes and glitches~\cite{Warszawski:2012zq}, accretion in a binary system, or the rearrangement of the star following the conversion of a hadronic star into a quark star~\cite{Lugones:2002vj,Abdikamalov:2008df}. Furthermore, in the post-merger phase of a binary neutron star system, the violent dynamics of the merging process can leave behind a massive neutron star that strongly oscillates~\cite{Andersson:2010ufc, Bauswein:2015vxa}. Specifically, the fundamental quadrupolar fluid mode of the remnant is strongly excited and dominates the post-merger gravitational wave (GW) signal.

Neutron star oscillations can be studied using the quasi-normal mode formalism \cite{ThorneCampolattaro1967, Lindblom:1983ps, Detweiler:1985zz}.  Research conducted over the past four decades has demonstrated that the $f$-mode of non-radial oscillations of neutron stars can yield crucial insights into the internal structure of these objects. Specifically, the frequency of this mode is highly dependent on the EOS, thereby providing a valuable means of probing the dense matter in the core of neutron stars. Additionally, the $f$-mode is particularly significant as it can be more easily excited than other higher frequency modes and can be detected by current gravitational wave detectors. In this work we will focus on the influence of the slope, $L$, on the $f$-mode of neutron stars.

\section{Nuclear model}
\label{sec:nuclear_model}

\subsection{Quantum hadrodynamics}

To describe the nuclear interaction, we use here an extended version of the QHD~\cite{Serot_1992}, which includes the $\omega-\rho$ non-linear coupling~\cite{IUFSU,Rafa2011,dex19jpg}, as well the scalar-isovector $\delta$ meson~\cite{KUBIS1997,Liu2002,Lopes2014BJP}. { In this study we focus  on purely nucleonic matter.} The Lagrangian density in natural units reads: 
\begin{equation}
\begin{aligned}
\mathcal{L}_{QHD} = & \bar{\psi}_N \big[\gamma^\mu(\mbox{i}\partial_\mu  - g_{\omega}\omega_\mu   - g_{\rho} \tfrac{1}{2}\vec{\tau} \cdot \vec{\rho}_\mu)  \\
& - (M_N - g_{\sigma}\sigma - g_{\delta}\vec{\tau} \cdot \vec{\delta}) \big]\psi_N  -U(\sigma)    \\
&  + \tfrac{1}{2}(\partial_\mu \sigma \partial^\mu \sigma - m_s^2\sigma^2)  \\
& + \tfrac{1}{2}(\partial_\mu \vec{\delta} \cdot \partial^\mu \vec{\delta} - m_\delta^2\delta^2)  \\ 
& - \tfrac{1}{4}\Omega^{\mu \nu}\Omega_{\mu \nu} + \tfrac{1}{2} m_v^2 \omega_\mu \omega^\mu   \\
& + \Lambda_{\omega\rho}(g_{\rho}^2 \vec{\rho^\mu} \cdot \vec{\rho_\mu}) (g_{\omega}^2 \omega^\mu \omega_\mu)  \\
& + \tfrac{1}{2} m_\rho^2 \vec{\rho}_\mu \cdot \vec{\rho}^{ \; \mu} - \tfrac{1}{4}\bf{P}^{\mu \nu} \cdot \bf{P}_{\mu \nu} . \label{s1} 
\end{aligned} 
\end{equation}
Here, $\psi_N$ is the Dirac field of the nucleons, and $\sigma$, $\omega_\mu$, $\vec{\delta}$, and $\vec{\rho}_\mu$ are the mesonic fields. The $g$'s are the Yukawa coupling constants that simulate the strong interaction, $M_N$ is the nucleon mass, and $m_s$, $m_v$, $m_\delta$, and $m_\rho$ are the masses of the $\sigma$, $\omega$, $\delta$, and $\rho$ mesons, respectively. The self-interaction term $U(\sigma)$, introduced in Ref.~\cite{Boguta} to fix the incompressibility, is given by:
\begin{equation}
U(\sigma) =  \frac{\kappa M_N(g_{\sigma} \sigma)^3}{3} + \frac{\lambda(g_{\sigma}\sigma)^4}{4} \label{s2} .
\end{equation} 
The Pauli matrices are denoted by $\vec{\tau}$, { the antisymmetric mesonic  field strength tensors are given by their usual expressions: $\Omega_{\mu\nu} = \partial_\mu \omega_\nu - \partial_\nu \omega_\mu$, and $P_{\mu\nu} = \partial_\mu \vec{\rho}_\nu - \partial_\nu \vec{\rho}_\mu -g_{\rho}(\vec{\rho}_\mu \times \vec{\rho}_\nu)$.   The $\gamma^\mu$ are the Dirac matrices and $\Lambda_{\omega\rho}$ is a non-linear isoscalar-isovector mixing coupling that provides  a simple and efficient method of softening the symmetry energy without compromising the success of the model in reproducing well-determined ground-state observables~\cite{IUFSU}. 	} The detailed calculation of the EOS in the mean field approximation can be found in ~\cite{Serot_1992,IUFSU,Lopes2014BJP,Lopes2022ApJ,Miy2021} and the references therein.

The symmetry energy for symmetric nuclear matter is defined as~\cite{KUBIS1997}:
\begin{equation}
\begin{aligned}
S(n) = & \frac{n}{8} \left( \frac{g_\rho}{m_\rho^{*}} \right)^2 + \frac{k_f^2}{6 \, ( M_N^{*2} +k_f^2)^{1/2}} \\
&  - \left( \frac{g_\delta}{m_\delta} \right)^2  \frac{M_N^{*2} ~ n/2 }{ (M_N^{*2} + k_f^2)[1 + (g_\delta/m_\delta)^2  A(k_f)]}  \label{s3},
\end{aligned} 
\end{equation}
where $n$ is the baryon number density, $k_f$ is the Fermi momentum, $m_\rho^{*} = (m_\rho^2 + 2\Lambda_{\omega\rho}g_\rho^2g_\omega^2\omega_0^2)^{1/2}$ is the effective mass of the meson $\rho$, $M_N^{*} = M_N - g_\sigma \sigma_0 - g_\delta \tau_3 \delta_0$ is the nucleon effective mass (it is worth to emphasize that for symmetric nuclear matter, the $\delta$ field is zero), and $A(k_f)$ is given by:
\begin{equation}
    A(k_f) =  \frac{4}{(2\pi)^3} \int_0^{k_f} d^3k \frac{k^2}{(M_N^{*2} +k^2)^{3/2} } . 
\end{equation}

Finally, the slope of the symmetry energy at the saturation point, $L$, is defined as:
\begin{equation}
L = 3n\bigg (\frac{\partial S}{\partial n} \bigg ) \bigg |_{n =n_0}    . \label{s4}
\end{equation}
We can also define the slope for an arbitrary density, $L(n)$, as in Ref.~\cite{Roca2011}:
\begin{equation}
L(n) = 3n\bigg (\frac{\partial S}{\partial n} \bigg )    . \label{s5}
\end{equation}

\subsection{Parametrization and symmetric nuclear matter}

{In this work, we use { a slightly modified version of the} L3$\omega\rho$ parametrization proposed in Ref.~\cite{Lopes2022CTP}.
The model parameters and their predictions for symmetric nuclear matter are presented in Table \ref{TL1}. The nuclear constraints at saturation density, taken from two extensive review articles  \cite{Dutra2014, Micaela2017}, are also included in Table \ref{TL1}. We select the parameters $(g_\rho/m_\rho)^2$, $(g_\delta/m_\delta)^2$, and $\Lambda_{\omega\rho}$ in order to fix the symmetry energy at the saturation point $S_0$ = 31.7 MeV { (which is slightly higher than the original value, $S_0$ = 31.2 MeV~\cite{Lopes2022CTP})},  while varying the slope $L$. The values of these parameters are reported in Table~\ref{T2}.}

\begin{table}[tb]
\begin{center}
\begin{tabular}{cc|cccc}
\toprule
  & Parameters & &  Constraints  & Our model  \\
\toprule
 $\left(g_{\sigma}/{m_s}\right)^2$ & $12.108 \, \mathrm{fm}^2$ &$n_0 (\mathrm{fm}^{-3})$ & 0.148 - 0.170 & 0.156 \\
$\left(g_{\omega}/{m_v}\right)^2$  & $7.132 \, \mathrm{fm}^2$ & $M^{*}/M$ & 0.6 - 0.8 & 0.69  \\
  $\kappa$ & 0.004138 & $K \mathrm{(MeV)}$ & 220 - 260             &  256  \\
$\lambda$ &  -0.00390  & $S_0 \mathrm{(MeV)}$  & 28.6 - 34.4 &  31.7  \\
- &  - & $B/A \mathrm{(MeV)}$  & 15.8 - 16.5  & 16.2 \\
\toprule
\end{tabular}
\caption{Model parameters used in this study and their predictions for symmetric nuclear matter at saturation density. The phenomenological constraints were taken from Refs.~\cite{Dutra2014, Micaela2017}.}
\label{TL1}
\end{center}
\end{table}

\begin{center}
\begin{table}[bt]
\begin{center}
\begin{tabular}{ccccc}
\toprule
  $L$ (MeV) & $(g_\rho/m_\rho)^2$ ($\mathrm{fm}^2$) & $(g_\delta/m_\delta)^2$ ($\mathrm{fm}^2$)  & $\Lambda_{\omega\rho}$  \\
\toprule
 44 & 8.40 & 0  & 0.0515 \\
 60 & 6.16  & 0  & 0.0344 \\
 76& 4.90 & 0 & 0.0171        \\
92 &  4.06  & 0 & 0  \\
100 &  7.23 & 0.92 & 0   \\
108 &  10.41 & 1.85 & 0   \\
116 &  13.48 & 2.76 & 0   \\
\toprule
\end{tabular}
\caption{Model parameters selected to set the symmetry energy at $S_0$ = 31.7 MeV.} 
\label{T2}
\end{center}
\end{table}
\end{center}

\begin{figure}[ht]
  \begin{centering}
\begin{tabular}{c}
\includegraphics[width=0.333\textwidth,angle=270]{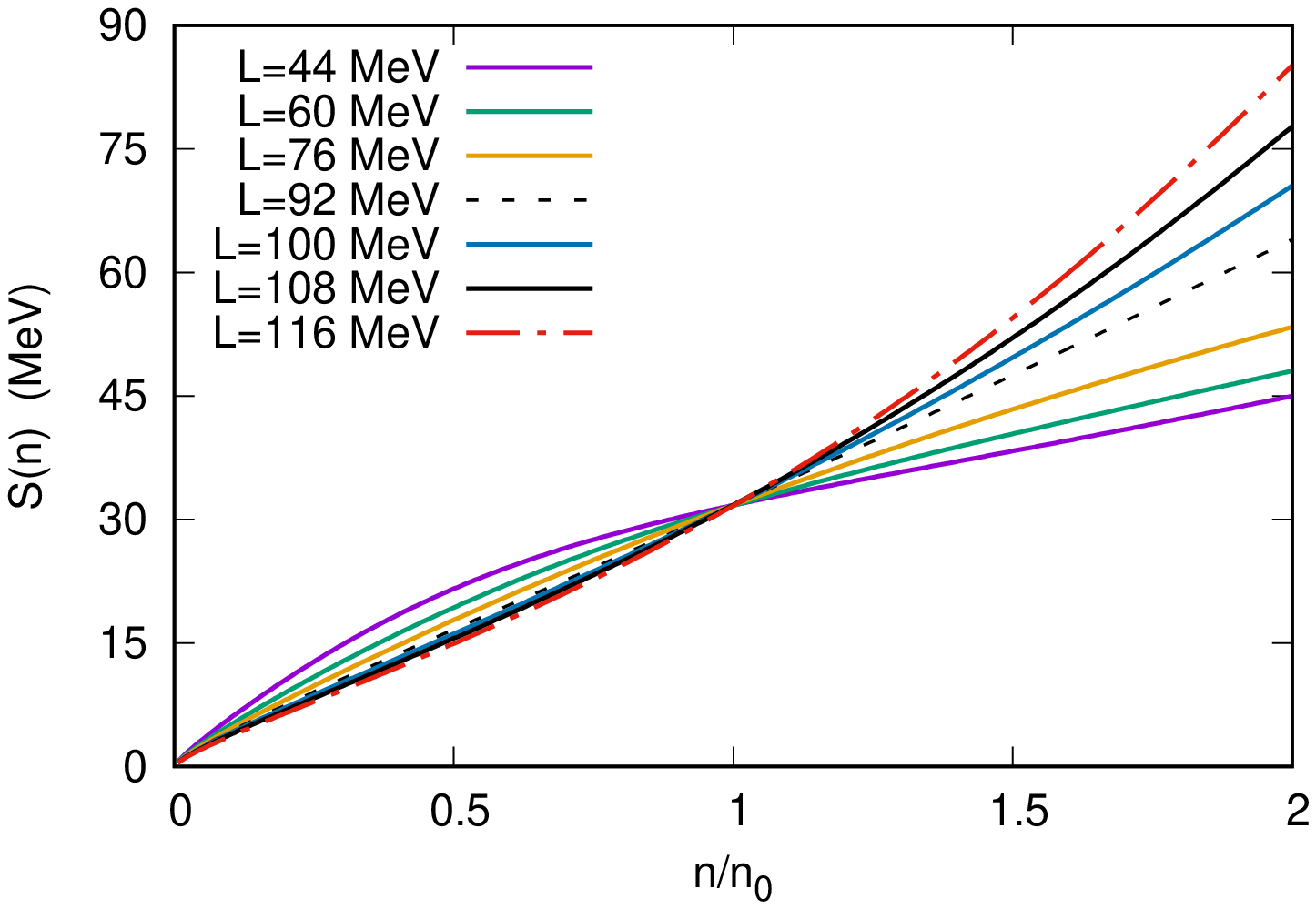} \\
\includegraphics[width=0.333\textwidth,,angle=270]{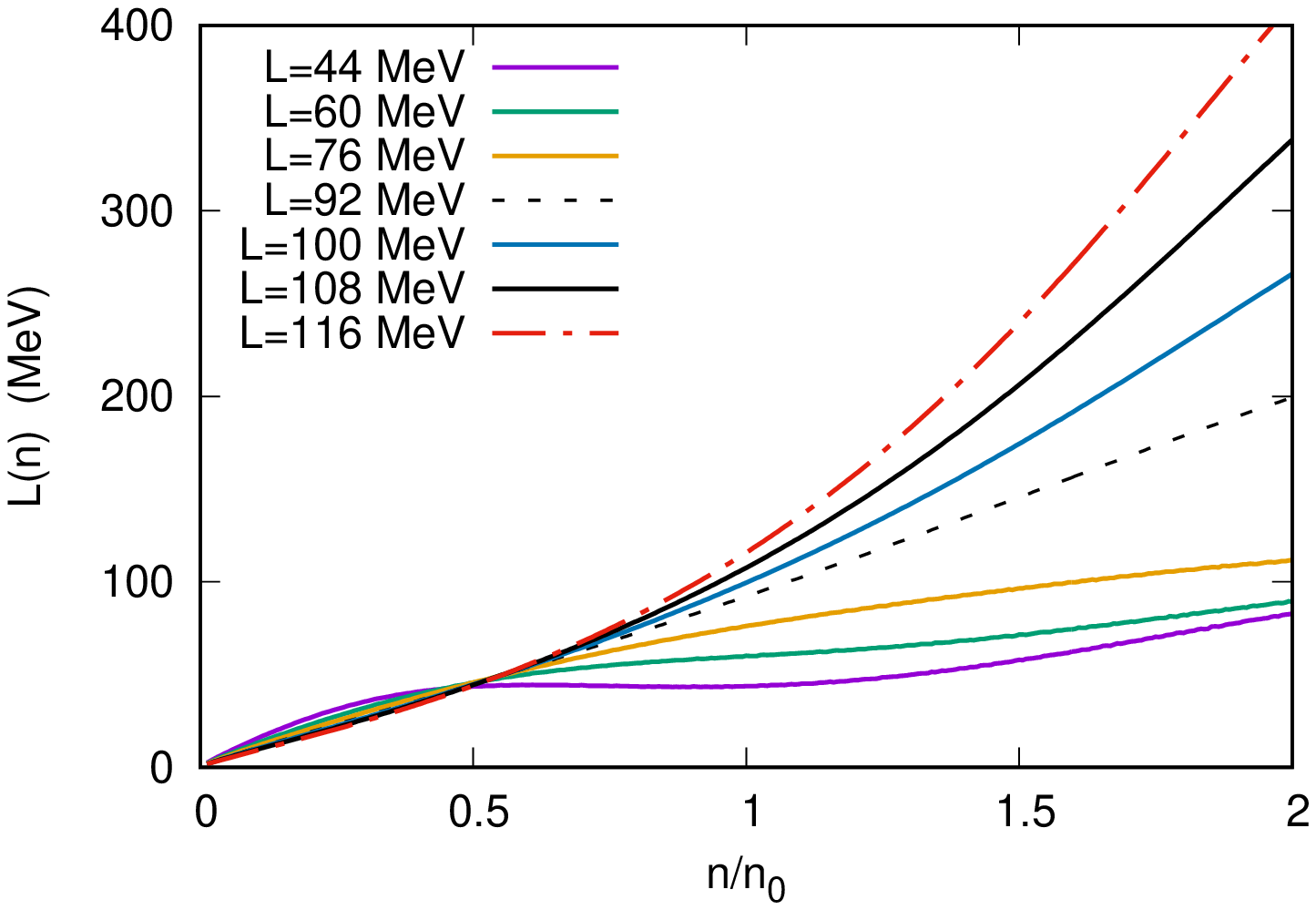} \\
\end{tabular}
\caption{Symmetry energy $S(n)$ and its slope $L(n)$ as a function of nucleon density up to twice the saturation density.} \label{F1}
\end{centering}
\end{figure}

Fig.~\ref{F1} shows the symmetry energy $S(n)$ and its slope $L(n)$ for densities up to twice the saturation density. We observe that the symmetry energy behaves oppositely for $n/n_0 < 1$ and $n/n_0 > 1$. Parametrizations with lower slopes have higher values of $S(n)$ for $n < n_0$ and lower values of $S(n)$ for $n > n_0$. This behavior is explained as follows. For parametrizations with nonlinear $\omega-\rho$ coupling, Table~\ref{T2} indicates that those with lower slopes have higher values of $(g_\rho/m_\rho)^2$ and $\Lambda_{\omega\rho}$. { The first term in Eq.~\eqref{s3} is more relevant at lower densities than at higher ones, owing to the fact that at lower densities, $m_\rho^{*}~\simeq~m_\rho$. As the density rises,  $m_\rho^*$ also increases (due to its dependence on the $\omega$ field, which in turn increases with the number density), consequently causing a decrease in $(g_\rho/m^*_\rho)^2$. This decline diminishes the impact of the first term in Eq.~\eqref{s3}.}

{ When we include the scalar-isovector $\delta$ meson, we observe that the contribution of the terms with the parameters $(g_\rho/m_\rho)^2$ and $(g_\delta/m_\delta)^2$ increase as the slope of the symmetry energy increases.} In this case, we have a competition between the repulsive vector $\rho$ meson and the attractive scalar $\delta$ meson. Since the vector field increases with the cube of the Fermi momentum, while the scalar field grows linearly~\cite{Serot_1992}, the attractive field dominates at lower densities, while the repulsive field dominates at higher densities.
The symmetry energy at twice the saturation density ranges from  $45 ~ \mathrm{MeV}$ for a slope of $L = 44 ~\mathrm{MeV}$ to  $85 ~\mathrm{MeV}$ for a slope of $L = 116 ~\mathrm{MeV}$.

For the slope at an arbitrary density, $L(n)$, we obtain analogous results { i.e., the values of lower $L_0$ present a larger value of $L(n)$ up to densities around $0.5n_0$, and then the results are reversed.}  We find that the slope at $n = 0.5n_0$ is approximately the same, $L(0.5n_0) \approx 44 ~\mathrm{MeV}$. However, at twice the saturation density, the slope varies from $L(2n_0) = 83 ~\mathrm{MeV}$ for $L = 44 ~ \mathrm{MeV}$ to $L(2n_0) = 412 ~ \mathrm{MeV}$ for $L = 116 ~\mathrm{MeV}$. { These behaviors can be explained by the nature of the couplings, analogous to the behavior of the symmetry energy.}

\section{Stellar structure and tidal deformability}

To investigate the effects of the nuclear symmetry energy slope on neutron star properties, we obtain the EOS of locally charge-neutral nuclear matter by including leptons (electrons and muons) in the system. Additionally, we assume that nuclear matter is in equilibrium under weak interactions and that the system is at zero temperature. To describe the outer crust and inner crust of the neutron star, we utilize the Baym-Pethick-Sutherland (BPS) EOS \cite{BPS} and the Baym-Bethe-Pethick (BBP) EOS \cite{BBP}, respectively.  { We use the BPS+BBP EoS up to the density of 0.0089 fm$^{-3}$ for all values of $L$, and from this point on, we use the QHD EoS, as suggested in Ref.~\cite{Glen}. In Ref.~\cite{Fortin2016}, the authors compare the BPS+BBP crust EoS with a unified EoS. They show that for the canonical star, there is a variation in the radius of $60 \mathrm{m} <~R_{1.4}~< 150 \mathrm{m}$. For a radius of 13 km, this implies an uncertainty around 1$\%$.} Once we have the crust+core EoS, the stellar structure is calculated by means of the Tolman-Oppenheimer-Volkoff (TOV) equations~\cite{TOV}:
\begin{eqnarray}
 \frac{dp}{dr} & = & - \frac{m \epsilon}{r^{2}} \bigg[ 1 + \frac{p}{\epsilon} \bigg ]  \bigg[ 1 + \frac{4\pi p r^3}{m} \bigg ] \bigg [ 1 - \frac{2m}{r} \bigg ]^{-1},  \\
 \frac{dm}{dr} & = & 4\pi r^2 \epsilon  . \label{stov}
\end{eqnarray}
{Here, $\epsilon$, $p$, and $m$ represent the energy density, pressure, and mass at the radial position $r$, respectively. We adopt units such that $G=c=1$.}

\begin{figure}[tb]
  \begin{centering}
\begin{tabular}{c}
\includegraphics[width=0.331\textwidth,angle=270]{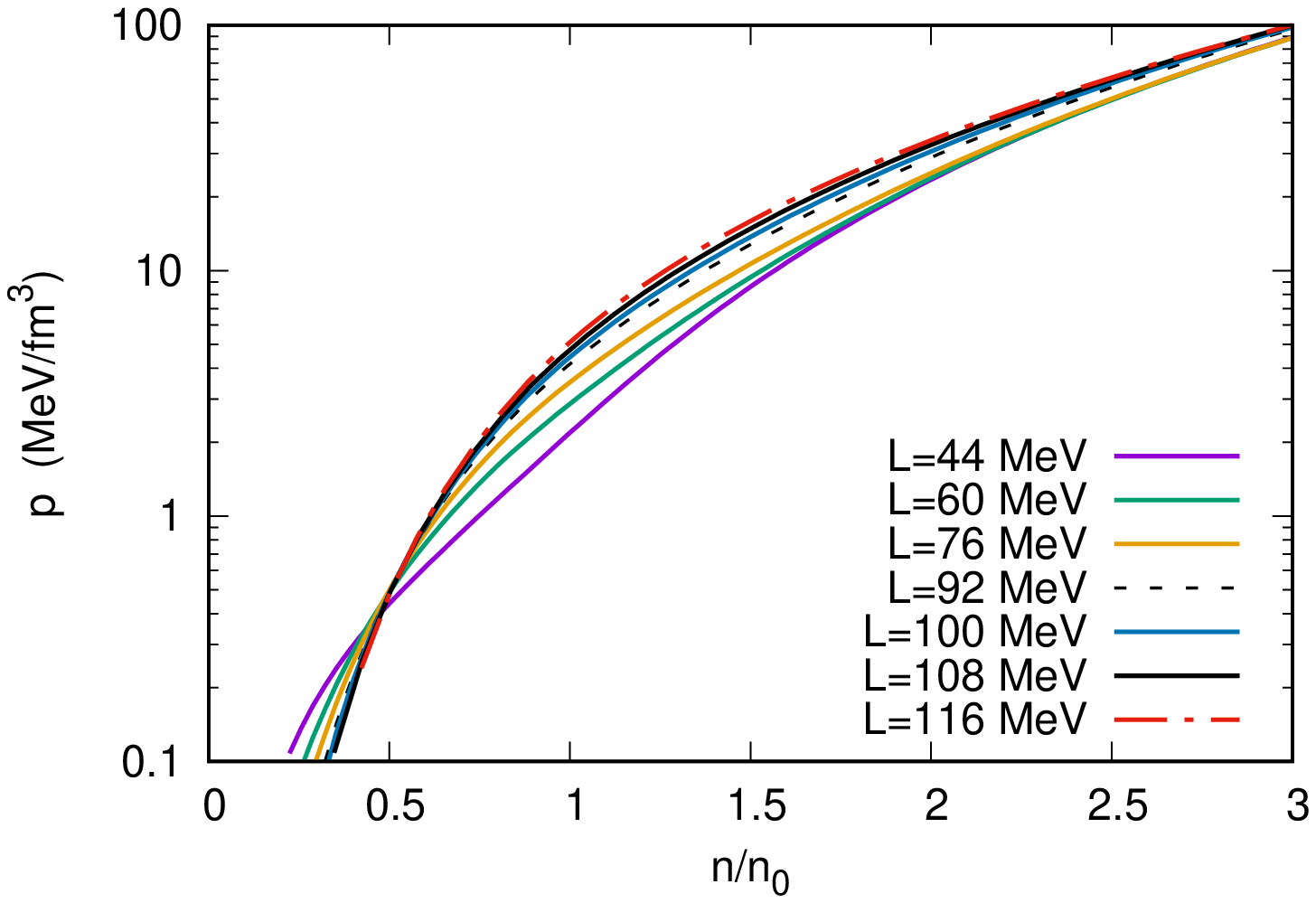} \\
\includegraphics[width=0.331\textwidth,angle=270]{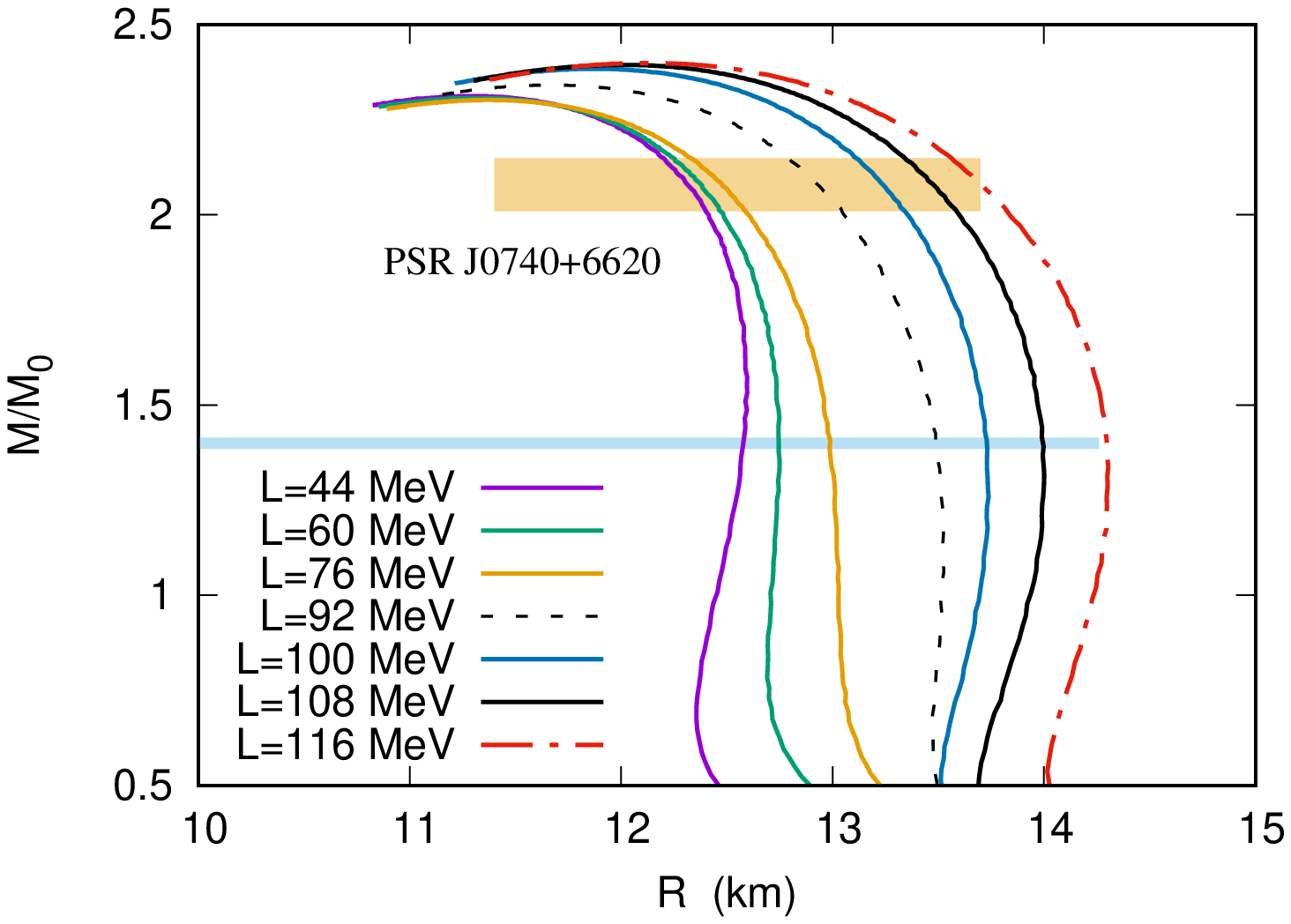} 
\end{tabular}
\caption{The figure depicts the EOS (upper panel) and mass-radius relationship (lower panel) for different $L$ values, along with observational constraints mentioned in the text. {In this figure, we take $M_0$ to represent the solar mass.}. The EOS curves converge at approximately at $0.5 n_0$, where all models have the same $L$ (see Fig. \ref{F1}). Beyond this point, a larger $L$ results in a stiffer EOS.  } \label{F2}
\end{centering}
\end{figure}

Fig.~\ref{F2} displays the EOS and the TOV solutions for different values of $L$. Despite the EOS curves are  beta-stable, while the slope $L(n)$ curves are relative to symmetric matter, both present similar behavior; i.e., the EOS is stiffer for low values of $L$ up to $n~\simeq$ 0.5 $n_0$. From this point on, this feature is reversed, and low values of $L$ present softer EOS. { This can be understood by expanding the EoS of asymmetric matter in terms of its symmetric components. Various theoretical studies (see, for instance Ref.~\cite{Li2008}) have shown that the energy per nucleon, $E$, of  asymmetric nuclear matter can be well approximated by:
\begin{equation}
E(n, \alpha)= E(n, \alpha=0) + S(n) \alpha^{2}+  O(\alpha^{4})    
\end{equation}
in terms of the baryon number density $n=n_{n}+n_{p}$, the isospin asymmetry $\alpha=(n_n-n_p) /(n_n + n_p)$, the energy per nucleon in symmetric nuclear matter $E(n, \alpha=0)$, and the bulk nuclear symmetry energy $S(n)$. 
The pressure is given by $p = n^{2} \frac{\partial(\epsilon / n)}{\partial n}$, and the energy per nucleon is related to the energy density as $E = \epsilon/n$.  Therefore: 
\begin{equation}
p(n, \alpha)  =  p(n, \alpha=0) + \frac{n}{3} \alpha^2  L(n) + n^2 S(n) \frac{\partial (\alpha^2)}{\partial n} + \cdots
\end{equation}
where $p(n, \alpha=0)$ is the pressure of symmetric matter. Neglecting the variation of $\alpha^2$ with $n$, we obtain that the lowest order correction to $p(n, \alpha=0)$ is directly related to $L(n)$: 
\begin{equation}
p(n, \alpha)  \approx  p(n, \alpha=0) + \frac{n}{3} \alpha^2  L(n) .
\end{equation}
Given that the first term is the same for all choices of $L$, the latter equation shows clearly why the slope $L(n)$ curves relative to symmetric matter present similar behavior as the pressure of beta stable matter around $n = 0.5n_0$   } However, as the inner-outer core transition, around $2 n_0$, is approached, the curves start to converge and exhibit similar levels of stiffness { since the nuclear interaction at high density is dominated by the $\omega$ meson}. The correlation between $L$ and the radii of NSs (i.e., that models with larger $L$ have larger radii) has been well established in the literature for the $\omega-\rho$ coupling~\cite{Rafa2011,Bao2020PRC}. In our study, we show that this correlation can also be extended to cases where the $\delta$ meson is included.  
Regarding the maximum mass, it varies from $2.30  M_\odot$ up to  $2.39 M_\odot$. Additionally, we observe that parametrizations with smaller values of $L$ - and therefore with the $\omega-\rho$ coupling - exhibit slightly lower maximum masses compared to those with higher slope values - and therefore with the scalar-isovector $\delta$ coupling.

We present a comparison of our results with recent constraints on the properties of neutron stars in Fig.~\ref{F2}. One of the most significant constraints is the existence of two solar masses NSs, which has been well-established by observations of pulsars such as PSR J1614-2230 \cite{Demorest:2010bx}, PSR J0348+0432 \cite{Antoniadis}, and PSR J0740+6620 \cite{Cromartie:2019kug}. Recent NICER observations have also constrained the radius of PSR J0740+6620 to be between $11.41 - 13.69 ~\mathrm{km}$ \cite{Riley2021}. Another, less stringent, constraint is the radius of the canonical $1.4 M_\odot$ star. While Ref. \cite{Ozel_2016} suggests an upper limit of only $11.1 ~\mathrm{km}$, Ref. \cite{Capano_2020} raises the limit to $11.9 ~\mathrm{km}$. Meanwhile, two NICER teams have pointed to a limit of $13.85 ~\mathrm{km}$ \cite{Riley:2019yda} and $14.26 ~\mathrm{km}$ \cite{Miller:2019cac} ($68\%$ credibility).  We adopt the larger of these values as a conservative limit on the radius of the canonical star. These constraints are represented by the hatched areas in Fig.~\ref{F2}. Notice that the constraint on the mass and radius of PSR J0740+6620 is satisfied for all values of $L$, although for $L = 116 ~\mathrm{MeV}$,  this constraint is close to the boundary of the 68\% credibility region. Concerning the radius of the canonical star, we observe that even though PREX2 implies that the slope could be as high as $143 ~\mathrm{MeV}$, a value of $L = 116 ~\mathrm{MeV}$ already leads to $R_{(1.4)}> 14.26 ~\mathrm{km}$.

\begin{figure}[tb]
  \begin{centering}
\begin{tabular}{c}
\includegraphics[width=0.31\textwidth,angle=270]{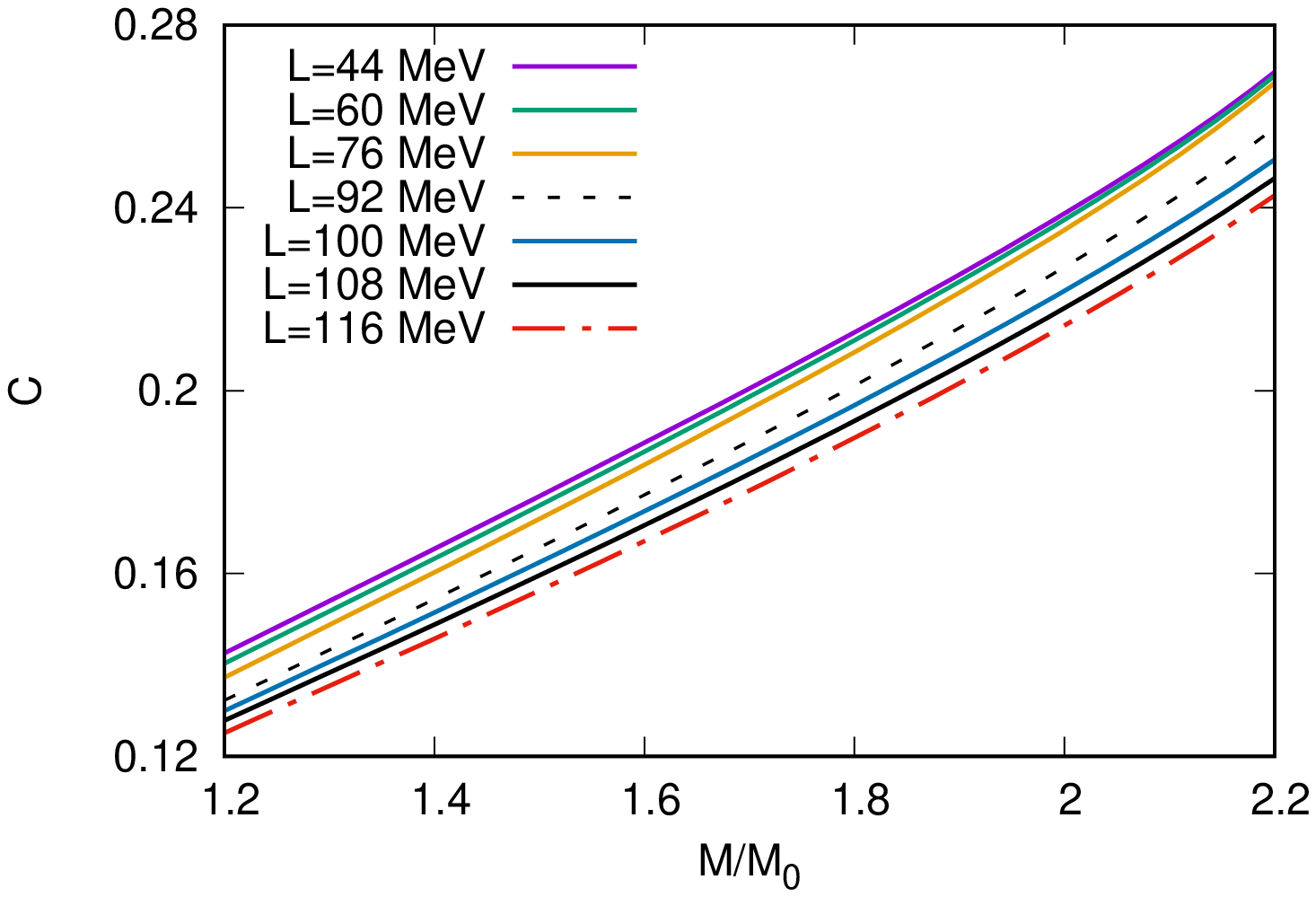} \\
\includegraphics[width=0.31\textwidth,,angle=270]{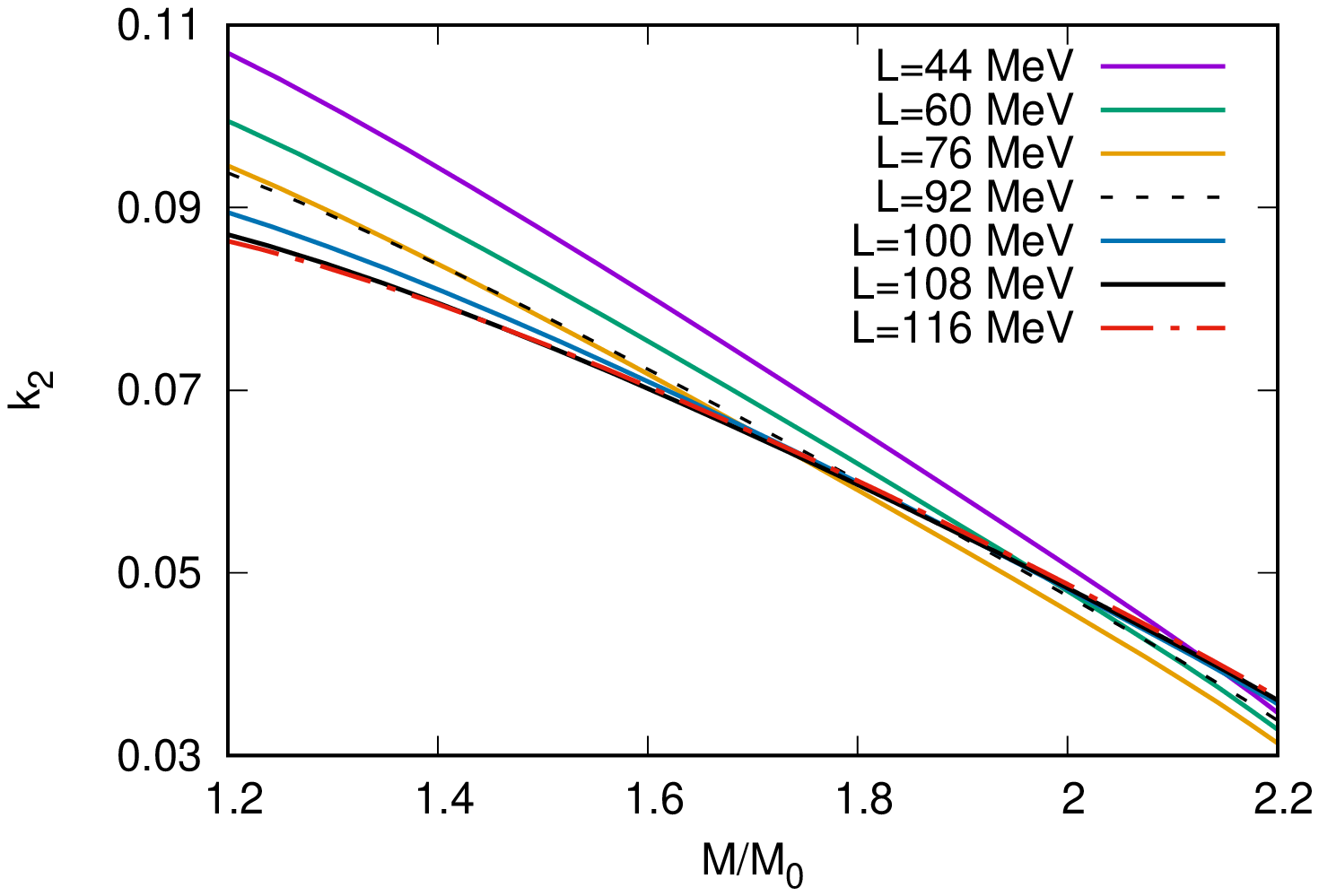} \\
\includegraphics[width=0.31\textwidth,,angle=270]{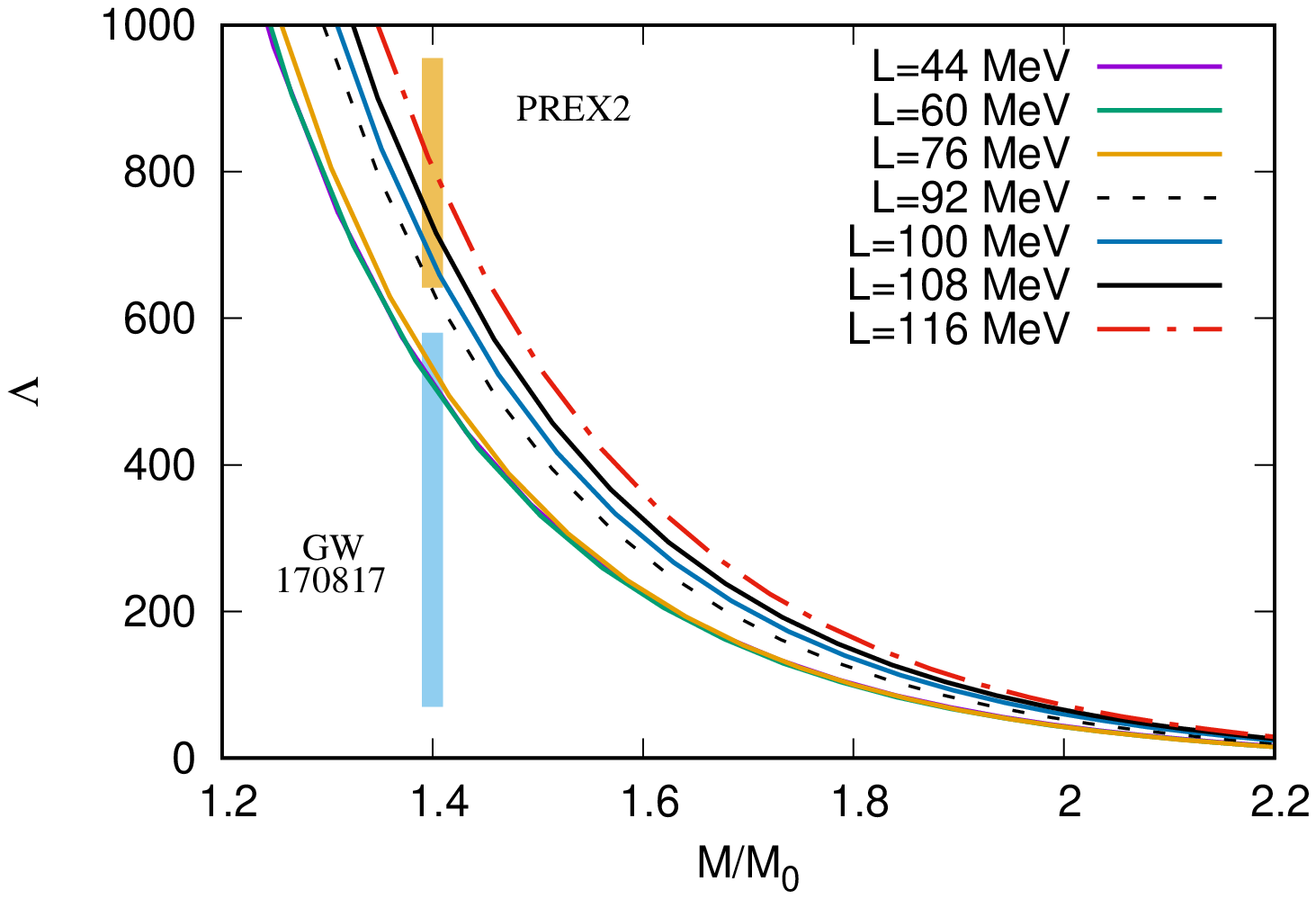} \\
\end{tabular}
\caption{The compactness (top), the second Love number (middle), and the dimensionless tidal parameter (bottom) as a function of the neutron star mass for different values of $L$.} \label{F3}
\end{centering}
\end{figure}

\begin{center}
\begin{table*}[tb]
\begin{center}
\begin{tabular}{c|cccccccccc}
\toprule
  $L (\mathrm{MeV})$ & $M_{\mathrm{max}} (M_\odot)$ & $R_{(1.4)}$ (km) & $C_{(1.4)}$ & $k_{2(1.4)}$ & $\Lambda_{(1.4)}$ & $R_{(2.01)}$ (km) & $C_{(2.01)}$ & $k_{2(2.01)}$ & $\Lambda_{(2.01)}$ \\
\toprule
 44 & 2.31 & 12.58  & 0.165 & 0.0947 & 515  & 12.40 & 0.240 & 0.0500 & 42\\
 60 & 2.30  & 12.74  & 0.163 & 0.0885 & 513 & 12.47 & 0.239 & 0.0473 & 40\\
 76& 2.30 & 12.99 & 0.160 & 0.0841   & 535 & 12.59 &  0.236 & 0.0452 & 40 \\
92 &  2.34  & 13.48 & 0.155 & 0.0839 & 639 & 13.04 & 0.228 & 0.0469 & 51 \\
100 &  2.37 & 13.73 & 0.151 & 0.0813 & 685 & 13.34 & 0.223 &  0.0476 & 58 \\
108 &  2.38 & 13.99 & 0.149 & 0.0799 & 728 & 13.59 & 0.219 & 0.0478 & 63 \\
116 &  2.39 & 14.30 & 0.146 & 0.0796 & 812 & 13.82 &  0.215 & 0.0482 & 70 \\
\toprule
\end{tabular}
\caption{For each value of the slope $L$, we give the maximum mass and the properties of the canonical $1.4 M_\odot$ star and the $2.01 M_\odot$ star.} 
\label{T3}
\end{center}
\end{table*}
\end{center}

{In recent years, the theory of relativistic tidal effects in binary systems has been the focus of intense research. Below, we  provide a summary of the procedure for computing the dimensionless tidal parameter $\Lambda$, which measures how easily an object is deformed by an external tidal field. It is defined as follows:
\begin{equation}
 \Lambda = \frac{2k_2}{3C^5} , \label{stidal}
\end{equation}
where $M$ is the mass of the compact object and $C = GM/R$ is its compactness. The parameter $k_2$ is known as the second-order Love number and is given by (\textbf{see Ref. \cite{Hinderer_2008}):}
\begin{equation}
\begin{aligned} 
k_{2}= & \frac{8 C^{5}}{5}(1-2 C)^{2}\left[2-y_{R}+2 C\left(y_{R}-1\right)\right] \\ & \times\left\{2 C\left[6-3 y_{R}+3 C\left(5 y_{R}-8\right)\right]\right. \\ & +4 C^{3}\left[13-11 y_{R}+C\left(3 y_{R}-2\right)+2 C^{2}\left(1+y_{R}\right)\right] \\ & \left.+3(1-2 C)^{2}\left[2-y_{R}+2 C\left(y_{R}-1\right)\right] \ln (1-2 C)\right\}^{-1}
\end{aligned}
\end{equation}
where $y_R=y(r=R)$ and $y(r)$ is obtained by solving:
\begin{equation}
 r\frac{dy}{dr} +y^2 + yF(r) +r^2Q(r) = 0 . \label{EL15}
\end{equation}
{The coefficients $F(r)$ and $Q(r)$ are given by:
\begin{eqnarray} 
F(r) &= & {\left[1-4 \pi r^{2}(\epsilon-p)\right]\left[1-\frac{2 m}{r}\right]^{-1} } , \\ 
Q(r) & = & 4 \pi\left[5 \epsilon+9 p+\frac{\epsilon+p}{c_{s}^{2}}-\frac{6}{4 \pi r^{2}}\right]\left[1-\frac{2 m}{r}\right]^{-1}   \nonumber \\ 
& & -\frac{4 m^{2}}{r^{4}}\left[1+\frac{4 \pi r^{3} p}{m}\right]^{2}\left[1-\frac{2 m}{r}\right]^{-2} ,
\label{EL17}
\end{eqnarray}
where $c_{s}^{2} \equiv d p / d \epsilon$ is the squared speed of sound. The boundary condition for Eq. \eqref{EL15}  at $r = 0$ is given by $y(0) = 2$. To obtain the tidal Love number, we use the EOS of Sec. \ref{sec:nuclear_model} and integrate the TOV equations along with Eq. \eqref{EL15}.
}

{In Fig.~\ref{F3} we  show $C$, $k_2$ and $\Lambda$ as a function of the neutron star mass. As expected, the lower the slope, the higher the star's compactness.  However,  the Love number $k_2$, also increases when we reduce the slope, which creates a competition between $C$ and $k_2$ in the determination of $\Lambda$, at least for stars with $M < 1.8 M_\odot$.  Nonetheless, it can be seen that $\Lambda$, which depends on both  $C$ and $k_2$, mostly decreases with decreasing slope.  
This trend is valid for both models, i.e., the one with the $\omega-\rho$ coupling and the one with the scalar-isovector $\delta$ meson coupling.}

{The tidal parameter of the canonical star, $\Lambda_{(1.4)}$, is also a useful tool to constrain $L$, but different restrictions are presented in the literature, and they are mutually exclusive. For instance, the analysis of the GW170817 event in Ref. \cite{AbbottPRL} constrained the dimensionless tidal parameter to be between $70$ and $580$, { which suggests a rather soft EoS.} Meanwhile,  in Ref. \cite{PREX2} the PREX2 experiment, { combined with results from the NICER X-ray telescope,} constrained the tidal parameter to be between $642$ and $955$, { which suggests a rather stiff EoS.} { (Nevertheless, it is worth mentioning that PREX2  and CREX results are still being debated in the community}~\cite{Tagami2022}).}  As observed, the parametrizations that incorporate the $\omega-\rho$ coupling, and hence have lower values of $L$, align with the results from Ref. \cite{AbbottPRL}. Conversely, the parametrizations that include the coupling with the scalar-isovector $\delta$ meson and therefore have higher values of $L$ align with Ref. \cite{PREX2}. For $L = 92$ MeV, the value of $\Lambda_{(1.4)} = 639$ falls into an ambiguous range.  }

{We will now specifically focus on the role of $L$ in two cases: the canonical $1.4 M_\odot$ star and the $2.01 M_\odot$ star. As previously emphasized, the radius of the canonical star increases with the slope $L$, ranging from 12.58 km to 14.30 km. The tidal parameter of the canonical star also increases with $L$, although the models with the lowest $L$ values (44 MeV and 60 MeV) are almost indistinguishable. 
{ When the parameter $L$ is larger, the EOS becomes stiffer. A stiffer EOS means that matter is more resistant to compression, leading to a larger radius for a given NS mass. With the mass of the NS held constant, the tidal deformability scales with the fifth power of the NS radius. Consequently, a stiffer EOS leads to a larger tidal deformability for a given mass value. This accounts for the positioning of the $\Lambda$ curves in the lower panel of Fig.~\ref{F3}; those corresponding to higher $L$ values are situated above those associated with lower $L$ values.} 
It is worth noting that, except for $L = 116$ MeV, all other parametrizations result in $R_{(1.4)} < $ 14.26 km, which agrees with \cite{Miller:2019cac,PREX2}, and $\Lambda_{(1.4)} < 800$, in agreement with Ref. \cite{Abbott2017}. Regarding the 2.01 $M_\odot$ neutron stars, we observe that their radii also increase with $L$ and fall in the range of 12.40 km to 13.82 km. However, the Love number $k_{2(2.01)}$ exhibits a more erratic behavior, in contrast to $k_{2(1.4)}$ which exhibits a monotonically decreasing trend with $L$. Finally, we have $40 < \Lambda_{(2.01)} < 70$. In Table \ref{T3}, we provide a summary of some of the main properties of neutron stars for various values of the slope $L$.}

\section{Neutron star oscillations}

In the near future, gravitational wave observatories are expected to detect pulsation modes of compact stars that are excited in binary mergers and newly born compact objects associated with core-collapse supernovae. Among the vast family of oscillation modes of compact stars, the quadrupolar ($\ell = 2$) fundamental ($f$) mode is crucial because it is highly probable that it will be significantly excited, thereby enabling its detection by the upcoming generation of instruments. Observations of $f$-mode oscillations can provide valuable information about the internal structure and composition of neutron stars. For example, the frequency and damping rate of the $f$-mode oscillation can be used to constrain the EOS of the dense matter in the star's core (see \cite{Andersson:1997rn, VasquezFlores:2018tjl, PhysRevC.95.025808, Flores:2013yqa} and references therein). Here, we will explore the possibility of constraining the nuclear symmetry energy slope from $f$-mode observations. 

Non-radial oscillations of neutron stars are analyzed using first-order perturbation theory within the framework of general relativity. {Inside the star, the static Schwarschild like metric is perturbed and it is given by }\cite{1983ApJS...53...73L,1985ApJ...292...12D}:    
\begin{equation}
\begin{aligned}
ds^2  = & -e^{\nu}(1+r^{\ell}H_0Y^{\ell}_{m}e^{i\omega t})dt^2  \\
&- 2i\omega r^{\ell+1}H_1Y^{\ell}_me^{i\omega t}dtdr +   \\
& + e^{\lambda}(1 - r^{\ell}H_0Y^{\ell}_{m}e^{i \omega t})dr^2 \\
& + r^2(1 - r^{\ell}KY^{\ell}_{m}e^{i \omega t})(d\theta^2 + \sin^2\theta d\phi^2) ,
\end{aligned}    
\end{equation}
where $Y_{l m}(\theta, \phi)$ are the spherical harmonics and the functions $H_{0}$, $H_{1}$, and $K$ depend only on $r$.
The Lagrangian 3-vector fluid displacement $\xi^{j}$ describes the small amplitude motion of the perturbed configuration and can be represented in terms of perturbation functions $W(r)$ and $V(r)$ as:
\begin{eqnarray}
\xi^{r} &=& r^{\ell-1}e^{-\lambda/2}WY^{\ell}_{m}e^{i\omega t}, \\
\xi^{\theta} &=& -r^{\ell - 2}V\partial_{\theta}Y^{\ell}_{m}e^{i\omega t}, \\
\xi^{\phi} &=& -r^{\ell}(r \sin \theta)^{-2}V\partial_{\phi}Y^{\ell}_{m}e^{i\omega t} . 
\end{eqnarray}
As shown in Refs. \cite{ThorneCampolattaro1967, 1983ApJS...53...73L,1985ApJ...292...12D}, perturbed Einstein’s equations inside the star ($0 < r < R$), are reduced to the following first-order linear system of differential equations: 
\begin{eqnarray}
H_1' &=&  - \frac{1}{r}  \biggl[ \ell+1+ \frac{2Me^{\lambda}}{r} +4\pi   r^2 e^{\lambda}(p-\epsilon) \biggr]H_1  \nonumber\\
&& +  \frac{e^{\lambda}}{r}  \left[ H_0 + K - 16\pi(\epsilon+p)V \right] ,      \label{osc_eq_1}  \\
 K' &=&   \frac{H_0}{r} + \frac{\ell(\ell+1)}{2r} H_1   - \left[\frac{(\ell+1)}{r} -\frac{\nu'}{2} \right] K  \nonumber \\
&& - 8\pi(\epsilon+p) e^{\lambda/2}r^{-1} W \:,  \label{osc_eq_2} \\
 W' &=&  - (\ell+1)r^{-1} W   + re^{\lambda/2} \bigg[ e^{-\nu/2} \gamma^{-1}p^{-1}X  \nonumber \\
&& - \ell(\ell+1)r^{-2} V + \tfrac{1}{2}H_0 + K \bigg] \:,  \label{osc_eq_3}\\
 X' &=&  - \ell r^{-1} X + \frac{(\epsilon+p)e^{\nu/2}}{2}  \Biggl[ \left( \frac{1}{r}   +\frac{\nu'}{2} \right)H_0  \nonumber \\
&& + \left(r\omega^2e^{-\nu} + \frac{\ell(\ell+1)}{2r}\right) H_1   + \left(\tfrac{3}{2}\nu' - \frac{1}{r} \right) K \nonumber  \\
&& - \ell(\ell+1)r^{-2}\nu' V  -  2 r^{-1}   \Biggl( 4\pi(\epsilon+p)e^{\lambda/2}  \nonumber \\
&&  + \omega^2e^{\lambda/2-\nu}
 - \frac{r^2}{2} \biggl(e^{-\lambda/2}r^{-2}\nu'\biggr)' \Biggr) W \Biggr]  \:,
\label{osc_eq_4}
\end{eqnarray}
where the prime denotes a derivative with respect to $r$, $\gamma$ is the adiabatic index 
\begin{equation}
    \gamma = \frac{(\epsilon + p)}{p}\frac{dp}{d\epsilon},
\end{equation}
$X$ is given by
\begin{equation}
\begin{aligned}
X = & ~ \omega^2(\epsilon+p)e^{-\nu/2}V - \frac{p'}{r}e^{(\nu-\lambda)/2}W   \\ 
& + \frac{1}{2}(\epsilon+p)e^{\nu/2}H_0 ,
\end{aligned}    
\end{equation}
and $H_{0}$ fulfills the algebraic expression 
\begin{equation}
\begin{aligned}
\left[ 3M + \tfrac{(\ell+2)(\ell-1)}{2}r + 4\pi r^{3}p\right] H_{0} =  8\pi r^{3}e^{-\nu /2}X  \\
- \left[ \tfrac{1}{2}\ell(\ell+1)(M+4\pi r^{3}p)-\omega^2  r^{3}e^{-(\lambda+\nu)}\right] H_{1} \\
+ \big[ \tfrac{1}{2}(\ell+2)(\ell-1)r - \omega^{2} r^{3}e^{-\nu}    \\
-   r^{-1}e^{\lambda}(M+4\pi r^{3}p)(3M - r + 4\pi r^{3}p)\big] K.
\end{aligned}    
\end{equation}

The fluid quantities outside the neutron star vanish, and as a result {the perturbation equations of the vacuum Schwarschild metric reduce to the Zerilli equation:}
\begin{equation}
\frac{d^{2}Z}{dr^{*2}}=[V_{Z}(r^{*})-\omega^{2}]Z,
\end{equation}
where the Zerilli function $Z(r^{*})$ and its derivative $dZ(r^{*})/dr^{*}$  can be expressed as functions of the metric perturbations  $H_{0}(r)$ and $K(r)$ using the transformation provided in Eqs. (A27)$-$(A34) of Ref. \cite{1983ApJS...53...73L} (notice that there is a typographical error in Eq. (A29) of \cite{1983ApJS...53...73L}, see e.g. \cite{Tonetto:2020bie}).  In the equations above, $Z(r^{*})$  depends on the ``tortoise'' coordinate, which is given by $r^{*} = r + 2 M \ln (r/ (2M) -1)$ and  the effective potential  $V_{Z}(r^{*})$ is:
\begin{eqnarray}
V_{Z}(r^{*}) = \frac{(1-2M/r)}{r^{3}(nr + 3M)^{2}}[2n^{2}(n+1)r^{3} \nonumber \\
+ 6n^{2}Mr^{2}+18nM^{2}r + 18M^{3}],
\end{eqnarray}
where $n= (\ell-1) (\ell+2) / 2$.

The physical solution of the oscillation equations must satisfy the following boundary conditions: (a) The functions describing the oscillatory behavior of the star must be regular at $r=0$. (b) The Lagrangian perturbation in pressure must be zero at the surface of the star, which implies that $X$ must  vanish at $r = R$. For  given values for $\ell$ and $\omega$, there exists a unique solution that satisfies the above boundary conditions inside the star. (c) Finally, at $r=\infty$, the physical solution of the Zerilli equation must describe purely outgoing gravitational radiation.

\begin{figure}[tb]
  \begin{centering}
\begin{tabular}{c}
\includegraphics[width=0.333\textwidth,angle=270]{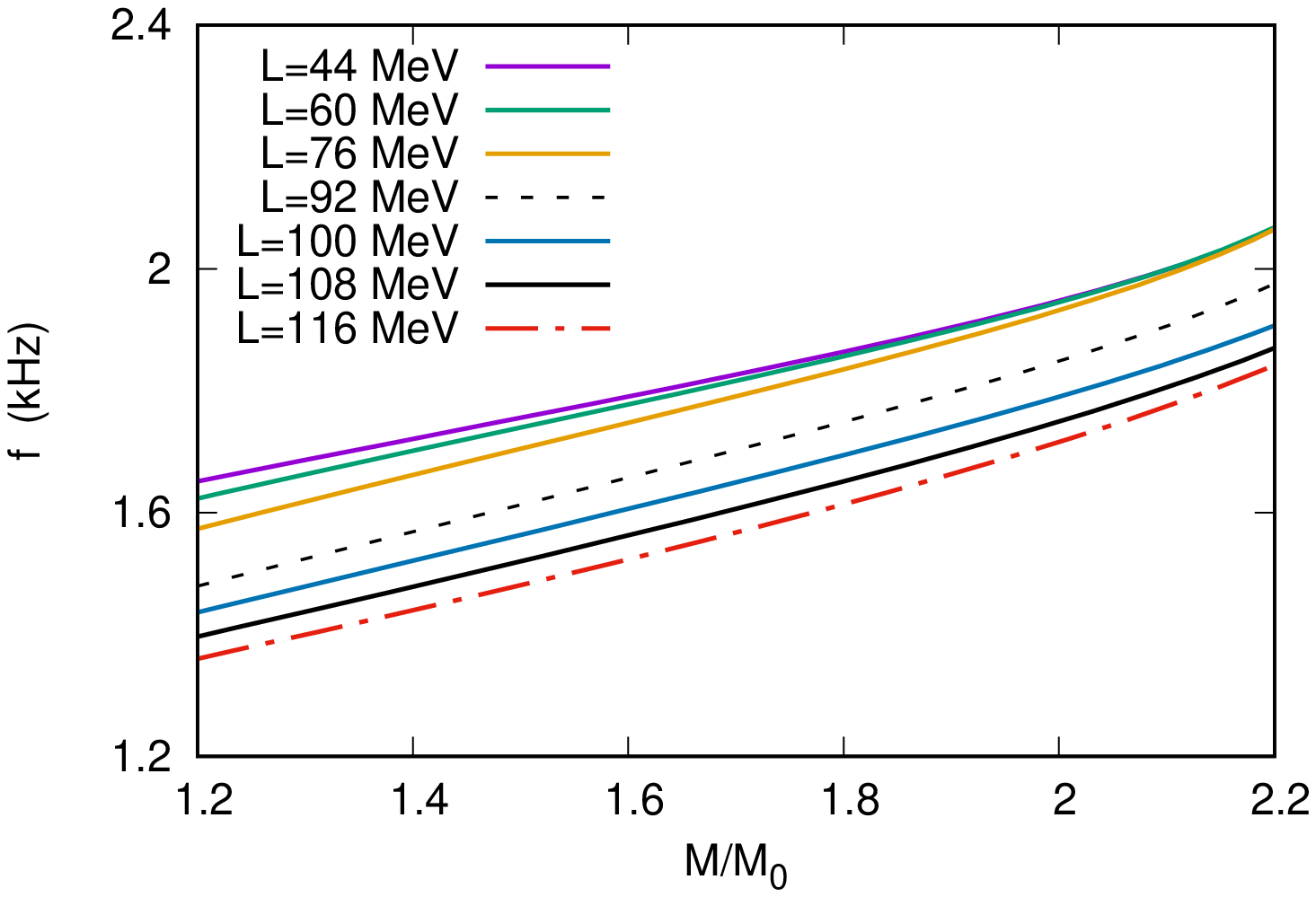} \\
\includegraphics[width=0.333\textwidth,,angle=270]{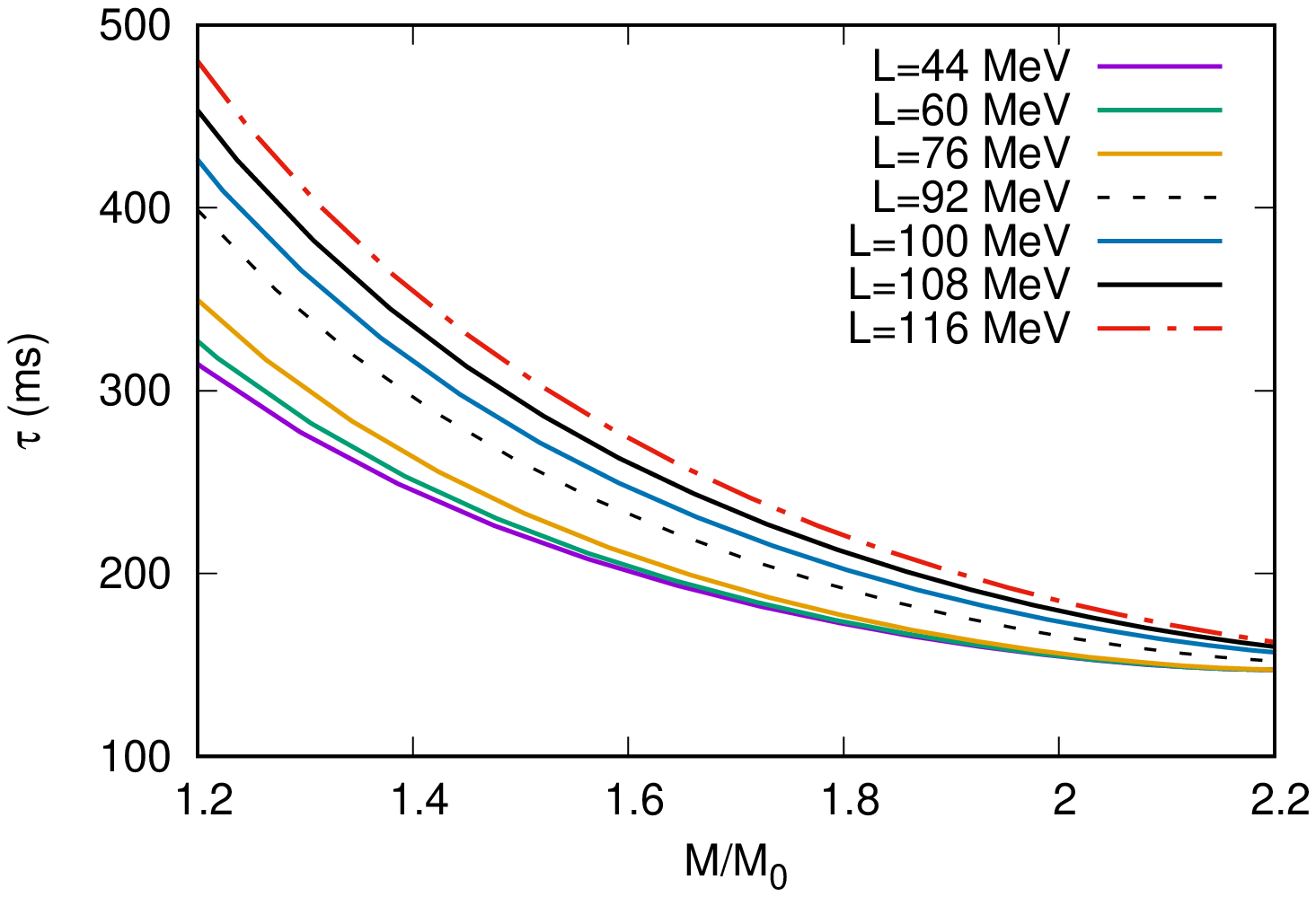} \\
\end{tabular}
\caption{Frequency (top) and damping  time (bottom) as a function of the neutron star mass for different values of $L$.} \label{F4}
\end{centering}
\end{figure}

The $f$-mode frequency and damping time are shown in Fig. \ref{F4} as a function of the stellar mass for different values of the nuclear symmetry energy slope $L$. The top panel shows that an increase in $L$  causes a systematic decrease in the gravitational wave frequencies. This behavior can be explained by the stiffening of the EOS as the value of $L$ increases. If a star has a stiffer EOS, its radius tends to be larger for a given mass, which in turn decreases its average density. Consequently, since the frequency of the $f$-mode scales with $\sqrt{M/R^3}$, the stiffer the EOS, the smaller the oscillation frequency, as demonstrated in Fig. 2 of Ref. \cite{VasquezFlores:2018tjl}. { A complementary  effect was noted in Ref.~\cite{Bikram2021}. In this work, the authors also found that a stiffer EOS results in a lower  $f$-mode frequency by varying the effective nucleon mass.} In our work, this effect is noticeable for all the parameters examined. However, for masses greater than $1.8 M_\odot$ and for $L \leq 76 ~ \mathrm{MeV}$, this effect is relatively small. The bottom panel shows that an increase in the value of  $L$ results in an increase in the damping time. This effect is more pronounced for small masses and becomes less noticeable for masses above $2 M_\odot$. In Table \ref{T4}, we present the $f$-mode frequencies and damping times for stars with masses of 1.4$M_\odot$ and 2.01$M_\odot$. For both mass values, an increase in $L$  results in a decrease in the $f$-mode frequency and an increase in its damping time. However, as already discussed in Fig. \ref{F4}, the effect of $L$ is less pronounced for more massive stars compared to less massive ones.

\begin{center}
\begin{table}[tb]
\begin{center}
\begin{tabular}{cccccc}
\toprule
  $L (\mathrm{MeV})$  & $f_{(1.4)} (\mathrm{kHz})$ & $\tau_{(1.4)}(\mathrm{ms})$   & $f_{(2.01)} (\mathrm{kHz})$  & $\tau_{(2.01)} (\mathrm{ms})$  \\
\toprule
 44 & 1.72 & 245  & 1.95 & 154 \\
 60 & 1.70  & 251  & 1.95 & 154 \\
 76& 1.66 & 264 & 1.94   & 156     \\
92 &  1.57  & 297 & 1.85 & 165  \\
100 & 1.52  & 316 & 1.80 & 173  \\
108 &  1.48 & 335 & 1.76 & 178  \\
116 & 1.44  & 354 & 1.72  & 184 \\
\toprule 
\end{tabular}
\caption{Frequency and damping time for the  $1.4 M_\odot$ and  $ 2.01 M_\odot$ stars.} 
\label{T4}
\end{center}
\end{table}
\end{center}

\section{Summary and Conclusions}

In this work we investigated the impact of the symmetry energy slope $L$ on various properties of neutron stars employing QHD with $\sigma-\omega-\rho-\delta$ mesons. We maintained a fixed value of symmetry energy at $n=n_0$ while varying the slope $L$ at $n_0$. Specifically, we decreased $L$ by introducing the non-linear $\omega-\rho$ coupling and increased it by including the scalar-isovector $\delta$ meson.  A stronger $\omega-\rho$ coupling led to a lower slope, whereas a stronger $\delta$ meson coupling resulted in a higher slope.  Parametrizations with higher slopes had lower values of the symmetry energy at $n~<n_0$ and higher values at $n>~n_0$. For $n \gtrsim 0.5 n_0$, the slope $L(n)$ follows a trend consistent with the slope $L$ defined at $n_0$. Specifically, as $L$ increases, so too does $L(n)$, as seen in Fig. \ref{F1}. A similar behavior is observed in the pressure of the beta-stable matter (Fig.~\ref{F2}). 

The maximum mass of a neutron star tends to increase slightly (less than 4\%) with an increase in the slope $L$. Meanwhile, the stellar radius is more sensitive to variations in $L$, increasing as $L$ increases. For instance, Table \ref{T3} shows that when $L$ rises from 44 MeV to 116 MeV, the radius $R_{(1.4)}$ of a canonical $1.4 M_\odot$ star  experiences an increase of approximately 15\%.
Our $M$-$R$ curves for low values of $L$  (i.e. in the range expected from neutron matter and nuclear binding energies, and also from measurements excluding PREX) are in excellent agreement with the 68\% credibility contour for the mass and radius of PSR J0030+0451 \cite{Miller:2019cac} and PSR J0740+ 6620 \cite{Miller2021}  and the 90\% credible level of the coalescing objects in the GW170817 event \cite{AbbottPRL}. 
However, as the value of $L$ increases, the $M$-$R$ curves shift towards larger radii. At around $L \sim 116 \, \mathrm{MeV}$,  the radius of the $1.4 M_\odot$ star is outside the upper boundary  of the NICER constraint on the radius of PSR J0030+0451, which places an upper limit of $14.26 \, \mathrm{km}$ at the 68\% level \cite{Miller:2019cac}. 
When even larger values of $L$ are considered, specifically those falling within the upper segment of the range provided by PREX2, we find that the resulting $M$-$R$ curves exhibit some tension with the aforementioned astrophysical observations, specially with GW170817.

Our study also involved calculating the dimensionless tidal deformability $\Lambda$ across a range of values for $L$. Our results, depicted in Fig.~\ref{F3}, show that $\Lambda$ generally increases as the slope $L$ increases. It is worth noting that $\Lambda$ exhibits a significant sensitivity to variations in $L$. For instance, Table \ref{T3} shows that $\Lambda_{(1.4)}$ varies between $515-812 \, \mathrm{MeV}$ and $\Lambda_{(2.01)}$ varies between $40-70 \, \mathrm{MeV}$ as $L$ increases from 44 to 116 $\mathrm{MeV}$, corresponding to a variation in $\Lambda$ of approximately $60-70\%$. These results are especially promising in light of the upcoming LIGO-Virgo-KAGRA observing period O4 scheduled for 2023, which is expected to enable resolution of the dimensionless effective tidal deformability with seven times greater accuracy than was achieved during O2 \cite{Coupechoux:2023fqq}. If such precise determinations become available, they could potentially impose significant constraints on the parameter $L$.

Finally, we focused on the impact of $L$ on the frequency and damping time of the $f$-mode. We have demonstrated that augmenting the value of $L$ results in a consistent reduction in the frequency, which can be ascribed to the increased stiffness of the core's EOS. In fact, a higher stiffness results in a greater radius for a given mass, thereby leading to a decrease in the star's average density. Hence, given that the frequency of the $f$-mode scales with $\sqrt{M/R^3}$, the larger the $L$ (stiffer EOS), the lower the oscillation frequency. Although the frequency exhibits a significant sensitivity to variations in $L$  for $L \gtrsim 70~\mathrm{MeV}$, we note that for masses beyond $1.8 M_\odot$ and $L \leq 76~\mathrm{MeV}$, this effect is relatively minor. Furthermore,  our analysis demonstrated that an increase in $L$ resulted in an increase in the $f$-mode damping time, with this effect being more conspicuous for lower masses and higher $L$ values. Although challenging, the detection of $f$-mode signals is expected to become a reality in the near future, due to the continued improvements in the sensitivity of GW observatories. The ability to detect $f$-mode frequencies and damping times would provide an additional independent tool to limit the range of possible values for the slope $L$.

\vspace{2cm}

\section*{Acknowledgements}

Luiz. L. Lopes would like to thank Professor Debora Peres Menezes for the help with some Fortran codes. GL acknowledges the financial support of the Brazilian agencies CNPq (grant 316844/2021-7) and FAPESP (grants 2022/02341-9 and 2013/10559-5). César O. V. Flores acknowledges the financial support of the Brazilian agency CNPq (grant 304569/2022-4). L. L. L. was partially supported by CNPq (Universal
Grant No. 409029/2021-1).

\bibliography{ref}

\end{document}